\documentclass[conference]{IEEEtran}
\usepackage{epsfig,epsf}%,fullpage}
\usepackage{color}
\usepackage{bbold}
\usepackage{multirow}
\usepackage{latexsym}
\usepackage{amssymb}
\usepackage{amsmath}

\usepackage{amsfonts}
\usepackage[sort]{cite}
\usepackage{fullpage}
\usepackage{mathrsfs}
\usepackage{scalefnt}
\usepackage{extarrows}
\usepackage{setspace}
\usepackage{stfloats}
\usepackage{url}
\usepackage{hyphenat}
\usepackage{graphicx}
\usepackage[table]{xcolor}
\usepackage{bbm}
\usepackage{dsfont}
\usepackage{bm}
\usepackage{array}
\usepackage{wrapfig}
\usepackage[]{algpseudocode}
\usepackage{algorithm}

\usepackage{mathtools}
\allowdisplaybreaks[1]
\usepackage{epstopdf}
\newcommand\numberthis{\addtocounter{equation}{1}\tag{\theequation}}
\pdfpageattr {/Group << /S /Transparency /I true /CS /DeviceRGB>>}

\makeatletter
\newsavebox\myboxA
\newsavebox\myboxB
\newlength\mylenA
\newcommand*\xoverline[2][0.75]{%
    \sbox{\myboxA}{$\m@th#2$}%
    \setbox\myboxB\null% Phantom box
    \ht\myboxB=\ht\myboxA%
    \dp\myboxB=\dp\myboxA%
    \wd\myboxB=#1\wd\myboxA% Scale phantom
    \sbox\myboxB{$\m@th\overline{\copy\myboxB}$}%  Overlined phantom
    \setlength\mylenA{\the\wd\myboxA}%   calc width diff
    \addtolength\mylenA{-\the\wd\myboxB}%
    \ifdim\wd\myboxB<\wd\myboxA%
       \rlap{\hskip 0.5\mylenA\usebox\myboxB}{\usebox\myboxA}%
    \else
        \hskip -0.5\mylenA\rlap{\usebox\myboxA}{\hskip 0.5\mylenA\usebox\myboxB}%
    \fi}
\makeatother  
\usepackage{enumerate}
\newcommand\norm[1]{\left\lVert#1\right\rVert}

\def\phi{\varphi}

\def\tr{{\rm tr}}
\def\He{{\rm H}}

\def\cC{{\cal C}}

\def\bh{{\mathbf{h}}}

\def\bm{{\mathbf{m}}}

\def\bu{{\mathbf{u}}}
\def\bv{{\mathbf{v}}}
\def\bw{{\mathbf{w}}}
\def\bx{{\mathbf{x}}}
\def\by{{\mathbf{y}}}
\def\bz{{\mathbf{z}}}
\def\b0{{\mathbf{0}}}

\def\argmax{\mathop{\mathrm{argmax}}}

\def\bH{{\mathbf{H}}}
\def\bI{{\mathbf{I}}}

\def\cC{\mathcal{C}}

\def\cF{\mathcal{F}}
\def\cG{\mathcal{G}}
\def\cH{\mathcal{H}}

\def\cN{\mathcal{N}}

\def\cR{\mathcal{R}}

\def\cX{\mathcal{X}}
\def\cY{\mathcal{Y}}

\def \tr {\text{tr}}

\def \Re[#1]{\text{Re}\left(#1\right)}%{#1_R}
\def \Im[#1]{\text{Im}\left(#1\right)}%{#1_I}
\def \Cpx[#1]{\tilde{#1}}
\def \tx {x} % tx scalar case
\def \rx {y} % rx scalar case

\def \txv {\bx} % tx input vector
\def \rxv {\by} % rx output vector

\def \txm {X} % tx matrix
\def \rxm {Y} % rx matrix

\def \txA {\cX} % tx symbol alphabet
\def \rxA {\cY} % rx symbol alphabet
\def \chm {H} % channel matrix

\def \txt {\tx_{\rm t}}   %tx training scalar
\def \txmt {\txm_{\rm t}} %tx training matrix

\def \rxt {\rx_{\rm t}}   %rx training scalar
\def \rxmt {\rxm_{\rm t}} %rx training matrix

\def \txd {\tx_{\rm d}}   %tx data scalar
\def \txvd {\txv_{\rm d}} %tx data vector
\def \txmd {\txm_{\rm d}} %tx data matrix

\def \rxd {\rx_{\rm d}}   %rx data scalar
\def \rxvd {\rxv_{\rm d}} %rx data vector
\def \rxmd {\rxm_{\rm d}} %rx data matrix

\def \txvte[#1]{\txv_{\rm t,#1}} 
\def \rxvte[#1]{\rxv_{\rm t,#1}} 
\def \txvde[#1]{\txv_{\rm d,#1}} 
\def \rxvde[#1]{\rxv_{\rm d,#1}} 
\def \nvbe[#1]{\nv_{\rm b,#1}} 

\def \txde[#1]{\tx_{\rm d,#1}} 
\def \rxde[#1]{\rx_{\rm d,#1}} 

\def \txvta[#1]{\txv_{{\rm t},#1}} 
\def \rxvta[#1]{\rxv_{{\rm t},#1}} 
\def \txvda[#1]{\txv_{{\rm d},#1}} 
\def \rxvda[#1]{\rxv_{{\rm d},#1}} 
\def \nvba[#1]{\nv_{{\rm d},#1}} 

\def \txda[#1]{\tx_{{\rm d},#1}} 
\def \rxda[#1]{\rx_{{\rm d},#1}} 

\def \ptrve[#1]{\hat{p}_{(\txvte[#1],\rxvte[#1])}}

\def \snr {\rho}

\def \MuI {\bI} % mutual information
\def \Ent {\bH} % entropy
\def \Entfuntwo {\cH_2} %entropy function
\def \E {\mathbb{E}} % expectation
\def \chv {\bh} % channel vector
\def \chs {h} % channel scalar
\def \tn {M} % tx antenna numbers
\def \rn {N} % rx antenna numbers
\def \nv {\bv} % noise vector
\def \nm {V} % noise matrix

\def \nmt {\nm_{\rm t}}
\def \nmd {\nm_{\rm d}}

\def \Tb {{T_{b}}} %total blocklength
\def \Tt {T_{\rm t}} %block training number
\def \Td {T_{\rm d}} %block data number
\def \Ttopt {T_{\rm{t,opt}}}
\def \sign {\text{sign}}
\def \Ctr {C_{\rm t}}

\def \ratio {\alpha}
\def \Cavg {\cC} %average capacity
\def \qh {q_{\rm h}}
\def \qhhat {\hat{q}_{\rm h}}

\def \txvda {\txv_{\rm{d},a}}
\def \qtx {q_{\tx}}
\def \qtxhat {\hat{q}_{\tx}}

\def \rxvdhat {\hat{\rxv}_{\rm d}}

\def \snreff {\snr_{\rm {eff}}}

\def \betat {\beta_{\rm{t}}}

\def \rxvdhat {\hat{\rxv}_{\rm{d}}}

\def \qx {q_{\rm x}}
\def \qxhat {\hat{q}_{\rm x}}

\def \qxL {q_{\rm x,L}}
\def \qxLhat {\hat{q}_{\rm x,L}}
\def \qxO {q_{\rm x,O}}
\def \qxOhat {\hat{q}_{\rm x,O}}

\def \Reff {R_{\rm {eff}}}
\def \RavgCSIR {\cR_{\rm CSIR}}
\def \Ravgeff {\cR_{\rm eff}}
\def \RavgeffL {\cR_{\rm eff,L}}
\def \RavgeffO {\cR_{\rm eff,O}}
\def \CavgBus {\cC_{\rm Bussgang}}
\def \betatopt {\beta_{\rm {t,opt}}}
\def \betatoptL {\beta_{\rm {t,opt,L}}}
\def \betatoptO {\beta_{\rm {t,opt,O}}}
\def \nonlinear {f}
\def \gfunc {g} % describe the nonlinear probability

\begin{document}

\title{Training for Channel Estimation in Nonlinear Multi-Antenna Transceivers}
\author{\IEEEauthorblockN{Kang Gao, J. Nicholas Laneman, N. J. Estes, Jonathan Chisum, Bertrand Hochwald}\\
\IEEEauthorblockA{Department of Electrical Engineering, University of Notre Dame, Notre Dame, IN, 46556\\
Email: \texttt{\{kgao,jnl,nestes,jchisum,bhochwald\}@nd.edu}}}
\maketitle
\begin{abstract}
Recent efforts to obtain high data rates in wireless systems have focused on what can be achieved in systems that have nonlinear or coarsely quantized transceiver architectures.  Estimating the channel in such a system is challenging because the nonlinearities distort the channel estimation process.  It is therefore of interest to determine how much training is needed to estimate the channel sufficiently well so that the channel estimate can be used during data communication.  We provide a way to determine how much training is needed by deriving a lower bound on the achievable rate in a training-based scheme that can be computed and analyzed even when the number of antennas is very large.  This lower bound can be tight, especially at high SNR.  One conclusion is that the optimal number of training symbols may paradoxically be smaller than the number of transmitters for systems with coarsely-quantized transceivers.  We show how the training time can be strongly dependent on the number of receivers, and give an example where doubling the number of receivers reduces the training time by about 37 percent.

\end{abstract}

\IEEEpeerreviewmaketitle

\section{Introduction}
Nonlinear transceivers are being considered for high-frequency wireless communication because of their low cost and power advantages.  Examples, include systems with low-resolution (especially one-bit) analog-to-digital converters (ADCs) at the receiver \cite{nossek2006capacity,mollen2017uplink,li2017channel,li2016much,hong2018weighted,fan2015uplink,wen2016bayes} or digital-to-analog converters (DACs) at the transmitter \cite{saxena2016analysis,li2017downlink,jacobsson2017massive,kong2018nonlinear} or both \cite{usman2016mmse,gao2017power,gao2018beamforming,kong2018multipair,gao2018capacity}. Performance analysis has shown that high data rates with low error probability can be achieved with multiple antennas and channel state information (CSI) at the receiver \cite{fan2015uplink}. 

Training-based schemes are often used in practice to obtain channel information, where part of coherence interval is used for training and the rest for data. In \cite{hassibi2003much}, a lower bound of the capacity in a training-based scheme for a linear system is provided, and the optimal training length is analyzed by maximize the lower bound. Simple analysis using a worst-case noise analysis allows optimum training rules to be derived.

In \cite{li2016much,li2017channel}, lower bounds on the channel capacity with training-based schemes are provided for systems with linear transmitters and one-bit quantized receivers. The authors formulate the quantized output as the combination of signal, Gaussian noise, and quantization noise uncorrelated with the signal, and provide a lower bound by following \cite{hassibi2003much} and considering the worst-case additive noise that minimizes the input-output mutual information in a linear system at low SNR. We consider a more general model where either the transmitter or the receiver can have an arbitrary nonlinearity.

We consider a general system with nonlinear transceivers and provide a lower bound on the achievable rate (channel capacity). We do not attempt to approximate or linearize the transceiver architecture, but instead directly deal with the nonlinearity. A step-by-step method is provided to compute the lower bound. When we apply our bound to a system with linear transmitters and one-bit receivers, we can improve upon existing known training-based results. When we apply our bound to a system with one-bit transceivers, we find that the optimal number of training symbols can be smaller than the number of transmitters when we have more receivers than transmitters.  We find that the optimal training length can decrease strongly as the number of receivers increases.

\section{training-based scheme and capacity lower bound}
\label{sec:lower_bound}
We assume a classical discrete-time block-fading channel\cite{hassibi2003much}, where the channel is constant for some discrete time interval $\Tb$, after which it changes to an independent value that holds for another interval $\Tb$, and so on. We divide the interval into two phases: $\Tt$ for training and $\Td$ for data transmission, where $\Tt+\Td=\Tb$. For a general system with nonlinear transceivers, within one block of $\Tb$ symbols, the channel can be modeled by 
\begin{align}
\rxmt = \nonlinear\left(\sqrt{\frac{\snr}{\tn}}\chm\txmt+\nmt\right),\tr\txmt^{\He}\txmt=\tn\Tt,
\label{eq:training_model}
\\
\rxmd = \nonlinear\left(\sqrt{\frac{\snr}{\tn}}\chm\txmd+\nmd\right),\E\tr\txmd^{\He}\txmd=\tn\Td,
\label{eq:data_model}
\end{align}
where $\txmt\in\txA^{\tn\times\Tt}$ and $\txmd\in\txA^{\tn\times\Td}$ are matrices of  transmitted training signal and data signal, $\rxmt\in\rxA^{\rn\times\Tt}$ and $\rxmd\in\rxA^{\rn\times\Td}$ are the corresponding matrices of received signal, $\tn$ and $\rn$ are the number of transmitters and receivers, $\txA$ and $\rxA$ are the alphabet of the transmitted signal and received signal, $\chm$ is a random channel matrix, which is fixed in one coherent time interval, $\nmt$ and $\nmd$ are additive noise. Elements of $\nmt,\nmd$ are independent and identically distributed ({\it iid}) complex Gaussian $\cC\cN(0,\sigma^2)$. Elements of $\chm$ are {\it iid}\/ complex Gaussian $\cC\cN(0,1)$. $\snr$ is the expected signal power at the receiver. $\nonlinear(\cdot)$ is an elementwise nonlinear function which models the nonlinearity of each receiver.

The capacity per transmitter per channel-use is 
\begin{equation*}
\Ctr = \sup\limits_{p_{\txmd}(\cdot),\txmd\in\txA^{\tn\times\Td}} \frac{1}{\Tb\tn}\MuI(\txmt,\rxmt,\rxmd;\txmd),
\end{equation*}
where $\MuI(\cdot;\cdot)$ is the mutual information notation.  In general, this optimization is difficult to compute, especially for large $T$, $M$, and $N$.
We use a series of now-standard inequalities to obtain a tractable lower bound on this capacity.
Let $\txvd(k)$ and $\rxvd(k)$ be the $k$th column of $\txmd$ and $\rxmd$. A lower bound on $\Ctr$ can be obtained by considering $\txvd(k)$ to be {\it iid}\/ with some distribution $p_{\txvd}(\txv)$. Then, 
\begin{align*}
 &\Tb\tn\Ctr \geq \MuI(\txmd;\rxmd |\txmt,\rxmt) \\
 &= \Ent(\txmd|\txmt,\rxmt) - \Ent(\txmd|\txmt,\rxmt,\rxmd)\\
&= \sum_{k=1}^{\Td}\Ent(\txvd(k)|\txmt,\rxmt) - \sum_{k=1}^{\Td}\Ent(\txvd(k)|\txmt,\rxmt,\rxmd,\\
&\qquad\qquad\txvd(1),\cdots,\txvd(k-1))\\
&\geq \sum_{k=1}^{\Td}\left(\Ent(\txvd(k)|\txmt,\rxmt) - \Ent(\txvd(k)|\txmt,\rxmt,\rxvd(k))\right) \\
&=(\Tb-\Tt)\MuI(\txvd;\rxvd|\txmt,\rxmt).
\numberthis
\label{eq:lower_bound1}
\end{align*}
%for any $\ptxvd(\txv)$.

Then, a lower bound on $\Ctr$ is
\begin{equation}
    \Ctr\geq C_{\rm bound} =\max_{\Tt} \frac{\Tb-\Tt}{\Tb}\Reff(\Tt),
    \label{eq:lower_bound_capacity}
\end{equation}
where 
\begin{equation}
    \Reff(\Tt) = \frac{1}{\tn}\MuI(\txvd;\rxvd|\txmt,\rxmt)
\end{equation}
is the effective achievable rate per transmitter in each channel-use.
The corresponding training time that optimizes the lower bound is
\begin{equation}
    \Ttopt = \argmax_{\Tt}\frac{\Tb-\Tt}{\Tb}\Reff(\Tt).
    \label{eq:optimal_Tt_setup}
\end{equation}
The complexity of finding $\Reff(\Tt)$ is lower than $\Ctr$ since we do not need to find the optimizing $p_{\txmd}(\cdot)$.  Nevertheless, $\Reff(\Tt)$ is still non-trivial to find since the amount of averaging needed to compute $\MuI(\txvd;\rxvd|\txmt,\rxmt)$ is generally exponential in $\tn$ and $\rn$.

As a result, a tight approximation of $\Reff(\Tt)$ is needed that works for any $\tn$ and $\rn$.  Some efforts to make such an approximation include modeling nonlinear receivers as linear receivers with an equivalent extra noise, and then using a worst-case noise analysis to obtain a lower bound. For example, linear transmitters and coarsely quantized (one-bit) receivers are considered in \cite{li2017channel} and \cite{li2016much}, and a ``Bussgang decomposition" is applied to find an equivalent uncorrelated quantization noise in a worst-case noise analysis. We do not employ such methods, but compare the lower bound we obtain with those obtained with such methods.

We approximate the bound $C_{\rm bound}$ with its limit when $\tn,\rn,\Tb\to\infty$.  We then use a novel equivalent channel, that can be analyzed using the so-called ``replica method".  The replica method, a tool used in statistical mechanics  \cite{mezard1988spin}, has been applied in many communication system contexts \cite{tanaka2001analysis,tanaka2002statistical,tanaka2004statistical,guo2005randomly,guo2003multiuser,wen2016bayes}, neural networks \cite{gardner1988space,engel2001statistical}, error-correcting codes \cite{montanari2000statistical}, and image restoration \cite{nishimori1999statistical}. Although the replica method is used for a large scale limit, \cite{gao2018capacity} and \cite{moustakas2003mimo} also show that results obtained from the replica method provide a good approximation for small $\tn$ and $\rn$ ($\approx 8$). A mathematically rigorous justification of the replica method is elusive, but the success of the method maintains its popularity. 

Because we look at the large $\tn$ and $\rn$ limit, we define
\begin{equation}
    \ratio=\frac{\rn}{\tn},\qquad \beta=\frac{\Tb}{\tn},
\end{equation}
and (\ref{eq:lower_bound_capacity}) and (\ref{eq:optimal_Tt_setup}) become
\begin{equation}
\Cavg_{\rm{t}}\geq\Cavg_{\rm{bound}}=\max_{\betat}\frac{\beta-\betat}{\beta}\Ravgeff(\betat),\quad\betat=\frac{\Tt}{\Tb},
\label{eq:lower_bound_rate_large_M_N}
\end{equation}
\begin{equation}
\betatopt=\argmax_{\betat}\frac{\beta-\betat}{\beta}\Ravgeff(\betat),
\label{eq:optimal_Tt_setup_large_M_N}
\end{equation}
where 
\begin{equation}
  \Ravgeff(\betat)=\lim_{\tn\to\infty}\frac{1}{\tn}\MuI(\txvd;\rxvd|\txmt,\rxmt),
  \label{eq:Reff_RS_def}
\end{equation}
\begin{equation*}
    \Cavg_{\rm{bound}}= \lim_{\tn\to\infty}C_{\rm {bound}}, \quad \Cavg_{\rm{t}}=\lim_{\tn\to\infty}\Ctr.
\end{equation*}

We assume that the elements of the training matrix $\txmt$ are {\it iid}\/ with 0 mean and unit variance. We find an equivalent channel that has the same achievable rate per transmitter as $\Ravgeff$.  The following claim summarizes the method we use.

\textbf{Claim:} Consider a nonlinear system with an unknown channel, with training and data phase as shown in (\ref{eq:training_model}) and (\ref{eq:data_model}). Let the elements of $\txmt$ be {\it iid}\/ with zero mean and unit variance.  Let the elements of $\txmd$ also be {\it iid}\/ zero-mean unit-variance but not necessarily the same distribution as $\txmt$. As $\tn,\rn,\Tt\to\infty$, we have
\begin{equation}
    \Ravgeff(\betat)=\lim_{\tn\to\infty}\frac{1}{\tn}\MuI(\txvd;\rxvdhat|\hat{\chm})
    \label{eq:Reff_equivalence},
\end{equation}
where $\Ravgeff$ is defined in (\ref{eq:Reff_RS_def}), $\txvd,\rxvdhat$ are the input and output of a new channel
\begin{equation}
    \rxvdhat = \nonlinear\left(\sqrt{\frac{\snr}{\tn}}\hat{\chm}\txvd + \hat{\nv}_{\rm d}\right), 
    \label{eq:equivalent_noise_model}
\end{equation}
where the channel $\hat{\chm}$ is known by the receiver, the elements of $\hat{\chm}$ are {\it iid}\/ $\cC\cN(0,\qh)$, and the elements of $\hat{\nv}_{\rm d}$ are {\it iid}\/ $\cC\cN(0,\sigma^2+\snr(1-\qh))$,
and $1-\qh$ is the mean square error of the minimum mean square error (MMSE) estimator of $\chm$ based on the training phase:
\begin{equation}
    1-\qh=\lim_{\tn\to\infty}\frac{1}{\tn\rn}\E\left(\norm{\chm-\E(\chm|\txmt,\rxmt)}_{\rm{F}}^2\right),
    \label{eq:channel_MMSE_RS}
\end{equation}
where $\norm{\cdot}_{\rm{F}}$ is the notation of Frobenius norm. The $\Ravgeff$ depends on $\betat$ through $\qh$. 

The proof is shown in Appendix \ref{app:mutual_information_equivalence} and uses some common assumptions used for replica methods \cite{tanaka2001analysis,tanaka2002statistical,guo2005randomly,wen2016bayes}.  We omit these here.
%1) generalization from integer to real is valid, 2) two limits can commute, and 3) the replica symmetry assumption holds.

\section{Step-by-step computation of lower bound}

We partition the computational methods into small and large $\tn,\rn,\Tb$.  These ``recipes" are then applied to a particular one-bit transceiver architecture in Section \ref{sec:app}.

\begin{enumerate}
    \item Derive $\gfunc(z,\rx,\sigma^2_0)$ as a function of $z,\rx,\sigma^2_0$ from
    \begin{equation}
        \gfunc(z,\rx,\sigma^2_0)=P(\nonlinear(z+v)=\rx),\quad v\sim \cC\cN(0,\sigma^2_0),
        \label{eq:gfunc_discrete_out}
    \end{equation}
    when $\rxA$ is discrete, and from
    \begin{equation}
    \gfunc(z,\rx,\sigma^2_0)=\prod_{o\in\{\rm R,I\}}\frac{d}{dy}P([\nonlinear(z+v)]_o\leq [\rx]_o),
    \label{eq:gfunc_continuous_out}
\end{equation}
where $v\sim \cC\cN(0,\sigma^2_0)$, $[\cdot]_R$ and $[\cdot]_I$ are the real and imaginary part of the enclosed value, when $\rxA$ is continuous.
\end{enumerate}

Steps 2A) to 7A) below compute $C_{\rm bound}$ directly, while steps 2B) to 10B) compute $\Cavg_{\rm bound}$ defined in (\ref{eq:lower_bound_rate_large_M_N}). The former works for small $\tn$ and $\rn$ when the computation complexity is affordable. Otherwise, the latter is appropriate.  We now describe the former.

\subsection{Small $\tn,\rn,\Tb$}
In the following steps, we need to run steps from 2A) to 6A) for $\Tt=1,2,\cdots,\Tb$. Then, run step 7A) to compute $C_{\rm bound}$.

\begin{enumerate}[1{A})]
\setcounter{enumi}{1}
\item Derive
\begin{align*}
    &d_1(\txvd,\rxvd,\txmt,\rxmt,\Tt)=\prod_{k=1}^{\rn}\E_{\chv}\Bigg[\gfunc\big(\sqrt{\frac{\snr}{\tn}}\chv^T\txvd,\\
    &\rx_{{{\rm{d}},k}},\sigma^2\big)\prod_{p=1}^{\Tt}\gfunc\left(\sqrt{\frac{\snr}{\tn}}\chv^T\txv_{{\rm{t}}, p},\rx_{{\rm{t}},kp},\sigma^2\right)\Bigg],
\end{align*}
as a function of $\txvd,\rxvd,\txmt,\rxmt,\Tt$, where $\chv\sim\cC\cN(0,I)$. If both $\txA$ and $\rxA$ are discrete, the numerical values can be computed and stored. Otherwise, simplify the expression if possible. This process of storing or simplification is the same for the following steps.
\item Derive
\begin{align*}
    d_2(\txmt,\rxmt,\Tt)=\prod_{k=1}^{\rn}\E_{\chv}\prod_{p=1}^{\Tt}\gfunc\left(\sqrt{\frac{\snr}{\tn}}\chv^T\txv_{{\rm{t}}, p},\rx_{{\rm{t}},kp},\sigma^2\right)
\end{align*}
as a function of $\txmt,\rxmt,\Tt$, where $\chv\sim\cC\cN(0,I)$.
\item Derive
\begin{equation}
    d_3(\txvd,\rxvd,\txmt,\rxmt,\Tt)=\frac{d_1(\txvd,\rxvd,\txmt,\rxmt,\Tt)}{d_2(\txmt,\rxmt,\Tt)}
\end{equation}
as a function of $\txvd,\rxvd,\txmt,\rxmt,\Tt$.
\item Derive
\begin{align*}
    &d_4(\txmt,\rxmt,\Tt)=-\sum_{\txvd,\rxvd}\Big(p_{\txvd}(\txvd)d_3(\txvd,\rxvd,\txmt,\rxmt,\Tt)\\
    &\ln\frac{\sum_{\txvd}p_{\txvd}(\txvd)d_3(\txvd,\rxvd,\txmt,\rxmt,\Tt)}{d_3(\txvd,\rxvd,\txmt,\rxmt,\Tt)}\Big)
\end{align*}
as a function of $\txmt,\rxmt,\Tt$, where 
\begin{equation*}
    p_{\txvd}(\txvd) = \prod_{k=1}^{\tn}p_{\txd}(\tx_{{\rm d},k}),
\end{equation*}
$\tx_{{\rm d},k}$ is the $k$th element of $\txvd$. Here we used summation, but if $\txA,\rxA$ is continuous, the summation should be replaced with integral. This also applies to the following steps.

\item Compute $\Reff(\Tt)$ from
\begin{align*}
    \Reff(\Tt)=\frac{1}{\tn}\sum_{\txmt,\rxmt}p_{\txmt}(\txmt)d_2(\txmt,\rxmt,\Tt)d_4(\txmt,\rxmt,\Tt),
\end{align*}
where 
\begin{equation*}
    p_{\txmt}(\txmt)=\prod_{m=1}^{\tn}\prod_{p=0}^{\Tt}p_{\txt}(\tx_{{\rm t},mp}),
\end{equation*}
$\tx_{{\rm t},mp}$ is the $m$th row and the $p$th column of $\txmt$.

\item Compute $\Ttopt$ and $C_{\rm bound}$ from
\begin{equation*}
    \Ttopt = \argmax_{\Tt}\frac{\Tb-\Tt}{\Tb}\Reff(\Tt).
\end{equation*}
\begin{equation*}
C_{\rm bound} = \frac{\Tb-\Ttopt}{\Tb}\Reff(\Ttopt),
\end{equation*}
\end{enumerate}

Steps 2A) to 7A) come directly from standard information-theoretic arguments.  Unfortunately the computational complexity increases very rapidly with $\tn,\rn,\Tb$ because all of the dimensions of the quantities involved grow.  This technique therefore is limited to small values.

\subsection{Large $\tn,\rn,\Tb$}

We use the Claim above to show how a bound $C_{\rm bound}$ defined in (\ref{eq:lower_bound_capacity}) may be computed.
Discretize $\betat$ in the range $(0,\beta)$ to a target accuracy for optimization, denoted as $\beta_{{\rm t},k},k=1,2,\cdots$. Compute steps from 2B) to 9B) for every $k$. Step 10B) then computes $\Cavg_{\rm bound}$. 

\begin{enumerate}[1{B})]
\setcounter{enumi}{1}
\item  Derive
\begin{equation}
    \cG_1(n,q)=-\beta_{{\rm t},k}\ln\left(\sum_{\rx\in\rxA}\E_{\bu}\prod_{a=0}^{n}\gfunc(\sqrt{\snr}u_a,\rx,\sigma^2)\right),
    \label{eq:G1_def}
\end{equation}
as a function of $n,q$ in a form so that $\cG_1(n,q)$ is differentiable at $n=0$, where $\bu=[u_0,u_1,\cdots,u_n]^T$, the real and imaginary part of $\bu$ are $[\bu]_{\rm R}$ and $[\bu]_{\rm I}$, which are independent $\cN(0,\frac{1}{2}Q_{\rm h})$. $Q_{\rm h}$ is a $(n+1)\times (n+1)$ matrix with 1 as diagonal elements and $q$ as off-diagonal elements. This special structure of $Q_{\rm}$ enables us to simplify the expression to be differentiable at $n=0$ for any $\gfunc(\cdot,\cdot,\cdot)$ through a simple trick. The details of the trick can be found in our examples shown in next section. We consider $\rxA$ as discrete here. If $\rxA$ is continuous, we just need to change summation into integral. This also applies to the following steps.

\item Derive
\begin{equation}
    \cF_1(q,\hat{q})=\lim_{n\to 0}\frac{\partial}{\partial n}\cG_1(n,q)+q\hat{q}+\ln(1+\hat{q})-\hat{q}
    \label{eq:cF1_value}
\end{equation}
as a function of $q$ and $\hat{q}$.

\item Solve 
\begin{equation}
    \frac{\partial \cF_1}{\partial q}=0, \frac{\partial \cF_1}{\partial \hat{q}}=0,
    \label{eq:MSE_RS_solution}
\end{equation}
and get solution $q=\qh,\hat{q}=\qhhat$.  If there are multiple solutions, select the solution ($\qh,\qhhat$) that minimize $\cF_1(q,\hat{q})$. We only require numerical solution and numerical methods such as Newton's method and secant method can be applied to solve such equation efficiently \cite{cheney2012numerical}. When we have expectation over random variables  (often Gaussian) in the equation, we may need to Monte Carlo method. In many cases, only one or two Gaussian random variables is shown in the equation, which does not require high computation power to get an accurate solution. Now, with $\qh$, we are ready to derive $\Ravgeff$ according to (\ref{eq:Reff_equivalence}). $\qh$ depends on $\beta_{{\rm t},k}$, and the in the rest steps, only $\qh$ is needed.

\item Let $\hat{\snr}=\qh\snr,\hat{\sigma}^2=\sigma^2+(1-\qh)\snr$. Derive
\begin{equation}
    \cG_2(n,r)=-\alpha\ln\left(\sum_{\rx\in\rxA}\E_{\bw}\prod_{a=0}^{n}\gfunc(\sqrt{\hat{\snr}}w_a,\rx,\hat{\sigma}^2)\right)
    \label{eq:G2_def}
\end{equation}
as a function of $n$ and $r$ in a form so that $\cG_2(n,r)$ is differentiable at $n=0$, where $\bw=[w_0,w_1,\cdots,w_n]^T$, the real and imaginary part of $\bw$ are $[\bw]_{\rm R}$ and $[\bw]_{\rm I}$, which are independent $\cN(0,\frac{1}{2}Q_{\rm \tx})$, where $Q_{\rm \tx}$ is a $(n+1)\times (n+1)$ matrix with 1 as diagonal elements and $r$ as off-diagonal elements. This step is very similar to 2B).

\item Derive
\begin{align*}
    &\cG_3(n,\hat{r})=-\ln\E_{\tx_0,\cdots,\tx_n}\exp\Big(\sum_{0\leq a<b\leq n}\\
    &2\hat{r}\cdot\left([\tx_{a}]_{\rm R}[\tx_{b}]_{\rm R}+[\tx_{a}]_{\rm I}[\tx_{b}]_{\rm I}\right)\Big)
    \numberthis
    \label{eq:G3_def}
\end{align*}
as a function of $n$ and $\hat{r}$ in a form so that $\cG_3(n,\hat{r})$ is differentiable at $n=0$, where $[\tx_{a}]_{\rm R}$ and $[\tx_{a}]_{\rm I}$ are the real and imaginary part of $\tx_a$, and $\tx_a$ are {\it iid}\/ with distribution $p_{\rm \tx}(\tx_a)$. In many cases, the real and imaginary part of $\tx_a$ are {\it iid}\/ and it simplifies the expression by only considering the expectation on the real part. For continuous input (often Gaussian), the tricks we used on examples shown in the next section  can be applied to make the expression differentiable at $n=0$. For discrete input, the tricks used in \cite{tanaka2002statistical} can be applied. We also used the trick in our example, but the details are omitted.

\item Derive
\begin{equation}
    \cF_2(r,\hat{r})=\lim_{n\to0}\frac{\partial}{\partial  n}\left(\cG_2(n,r)+\cG_3(n,\hat{r})\right)+r\hat{r}
    \numberthis
    \label{eq:F2_def}
\end{equation}
as a function of $r$ and $\hat{r}$.

\item Solve 
\begin{equation}
    \frac{\partial \cF_2}{\partial r}=0, \frac{\partial \cF_2}{\partial \hat{r}}=0,
    \label{eq:qx_RS_solution}
\end{equation}
and get solution $r=\qx,\hat{r}=\qxhat$. If there are multiple solutions, select the solution ($\qx,\qxhat$) that minimize $\cF_2(r,\hat{r})$.

\item Compute $\Ravgeff(\beta_{\rm t,k})$ through
\begin{align*}
    \Ravgeff(\beta_{{\rm t},k}) &= \frac{1}{\ln 2}\Big[\cF_2(r,\hat{r}) + \ratio\sum_{\rx\in\rxA}\E_{u}\Big(
    \gfunc(\sqrt{\hat{\snr}}u,\rx,\hat{\sigma}^2)\\
    &\cdot \ln\gfunc(\sqrt{\hat{\snr}}u,\rx,\hat{\sigma}^2)\Big)\Big],u\sim\cC\cN(0,1)
    \numberthis
    \label{eq:Reff_RS_solution}
\end{align*}
Note that $\Ravgeff(\beta_{{\rm t},k})$ depends on $\beta_{{\rm t},k}$ through $\qh$.

\item Compute $\betatopt$  and $\Cavg_{\rm{bound}}$ through
\begin{equation*}
    \betatopt=\argmax_{\beta_{{\rm t},k}}\frac{\beta-\beta_{{\rm t},k}}{\beta}\Ravgeff(\beta_{{\rm t},k})
\end{equation*}
\begin{equation}
    \Cavg_{\rm{bound}}=\frac{\beta-\betatopt}{\beta}\Ravgeff(\betatopt)
    \label{eq:lower_bound_recipe}
\end{equation}

\end{enumerate}

The proof is shown in Appendix \ref{app:channel_estimation_MI_RS}.

\section{application of the lower bound}
\label{sec:app}
We consider two nonlinear systems as examples of the step-by-step methods presented in the previous section. We focus on large $\tn$ and $\rn$, and compute $\Cavg_{\rm bound}$. One system uses linear transmitters with $\txA$ arbitrary complex, and one-bit receivers with $\rxA=\{\pm1\pm j\}$.  The other system uses one-bit transmitters with $\txA=\{\frac{\pm1\pm j}{\sqrt{2}}\}$ and one-bit receivers with $\rxA=\{\pm1\pm j\}$. The nonlinear function is $\nonlinear(z)=\sign(z)$ for both cases, where the output of $\sign(z)$ is a complex number with the sign of the real and imaginary parts of $z$ as its real and imaginary parts.  This model mimicks having a highly-nonlinear single-bit quantizer in the transceiver chain.

We let $\sigma^2=1$, and therefore $\snr$ is the SNR at each receiver. For linear transmitters, we assume each element in $\txvd$ are {\it iid}\/ $\cC\cN(0,1)$, and for one-bit transmitters we assume each element in $\txvd$ are {\it iid}\/ uniform distributed in $\{\frac{\pm1 \pm j}{\sqrt{2}}\}$. With given $\tn,\rn,\Tt,\snr$, we treat $\betat$ as a variable and discretize it in increments of 0.1 for numerical accuracy.

For many of the steps, the distinction between linear transmitters and one-bit transmitters is not needed.  In the steps where the distinction is important, we use the subscripts ``L" and ``O" to indicate ``linear" or "one-bit" at the transmitter.

\begin{enumerate}
    \item Derive $\gfunc(z,\rx,\sigma^2_0)$ according to (\ref{eq:gfunc_discrete_out}), we have
    \begin{equation}
        \gfunc(z,\rx,\sigma^2_0)=\prod_{o\in\{\rm R,I\}}Q(-\frac{\sqrt{2}}{\sigma_0}[z]_o\cdot[\rx]_o),
    \end{equation}
    where $[\cdot]_{\rm R}$ and $[\cdot]_{\rm I}$ are the real and imaginary part of the enclosed value.
\end{enumerate}

\begin{enumerate}[1{B})]
\setcounter{enumi}{1}
\item Let $\ratio=\frac{\rn}{\tn}$. Derive $\cG_1(n,q)$ according to (\ref{eq:G1_def}):
\begin{align*}
    &\cG_1(n,q)   \\ &=-\betat\ln\prod_{o\in\{\rm{R,I}\}}\sum_{\rx\in\{\pm1\}}\E_{[\bu]_o}\prod_{a=0}^{n}Q(\sqrt{2\snr}[u_a]_o\rx)\\
    &=-2\betat\ln\left(2\E_{\bz}\prod_{a=0}^{n}Q(\sqrt{\snr}z_a)\right),
\end{align*}
where $\bz=[z_0,\cdots,z_n]^T\sim\cN(0,Q_{\rm \chs})$, the diagonal elements of $Q_{\rm \chs}$ are 1 and the off-diagonal elements of $Q_{\rm \chs}$ are $q$.

We can consider $z_a=\sqrt{q}u+\sqrt{1-q}t_a$ with $u,t_0,t_1,\cdots,t_n$ to be {\it iid}\/ $\cN(0,1)$. Then, we have
\begin{align*}
    &\cG_1(n,q)\\
    &= -2\betat\ln2\E_u[\E_{t_0}Q(\sqrt{\snr q}u + \sqrt{\snr(1-q)}t_0)]^{n+1}\\
    &= -2\betat\ln\left(2\E_u\left[Q(\sqrt{\frac{\snr q}{\snr(1-q)+1}}u)\right]^{n+1}\right).
    \numberthis
    \label{eq:cG1_solution_one_bit_receiver}
\end{align*}

\item Derive $\cF_1(q,\hat{q})$ according to (\ref{eq:cF1_value}):
\begin{align*}
    &\cF_1(q,\hat{q})=-4\betat\E_{u}\Big[Q\left(\sqrt{\frac{\snr q}{\snr(1-q)+1}}u\right)\\
    &\cdot \ln Q\left(\sqrt{\frac{\snr q}{\snr(1-q)+1}}u\right)\Big] +q\hat{q}+\ln(1+\hat{q})-\hat{q}.
\end{align*}

\item Solve $\qh$ according to (\ref{eq:MSE_RS_solution}):

By solving (\ref{eq:MSE_RS_solution}), we eventually get that $\qh$ is the solution of
\begin{equation}
    \frac{q}{1-q}=\frac{\betat B^2}{\pi}\E_{u}\left(\frac{\exp(-B^2qu^2)}{Q(B\sqrt{q}u)}\right),
    \label{eq:qh_solution_one_bit_rx}
\end{equation}
where $B=\sqrt{\frac{\snr}{1+\snr(1-q)}},u\sim\cN(0,1)$. The solution $\qh$ depends on $\betat$, and in the rest steps, only $\qh$ is needed.

\item Let $\hat{\snr}=\qh\snr,\hat{\sigma}^2=1+(1-\qh)\snr$. Derive $\cG_2(n,r)$ according to (\ref{eq:G2_def}):

$\cG_2(n,r)$ is very similar to $\cG_1(n,q)$. Similar to (\ref{eq:cG1_solution_one_bit_receiver}), we have
\begin{align*}
    &\cG_2(n,r)\\
    &= -2\alpha\ln\left(2\E_u\left[Q(\sqrt{\frac{\snreff r}{\snreff(1-r)+1}}u)\right]^{n+1}\right),
    \numberthis
    \label{eq:cG2_solution_one_bit_receiver}
\end{align*}
where
\begin{equation}
    \snreff=\frac{\snr\qh}{1+\snr(1-\qh)}.
    \label{eq:effective_SNR}
\end{equation}

\item Derive $\cG_3(n,\hat{r})$ according to (\ref{eq:G3_def}):

For linear transmitter, we consider each element of $\txvd$ are i.i.d $\cC\cN(0,1)$.
\begin{align*}
    \cG_{3,\rm L}(n,\hat{r}) = -2\ln\E_{\bw}\exp(\sum_{a<b}\hat{r}w_aw_b),
\end{align*}
where $\bw=[w_0,w_1,\cdots,w_n]^T\sim\cN(0,I)$.

Since
\begin{align*}
    &\E_{\bw}\exp(\sum_{a<b}\hat{r}w_aw_b)=\int_{\cR^{n+1}}\frac{d\bw}{(2\pi)^{\frac{n+1}{2}}}\exp(-\frac{1}{2}\bw^T D \bw)\\
    &=|D|^{-\frac{1}{2}},
\end{align*}
where the diagonal elements of $D$ are 1 and off-diagonal elements are $-\hat{r}$. Therefore, we have
\begin{equation*}
    \cG_{3,\rm L}(n,\hat{r}) = (n+1)\ln(1+\hat{r})+\ln\left(1-\frac{(n+1)\hat{r}}{1+\hat{r}}\right).
\end{equation*}

For one-bit transmitter, we consider each element of $\txd$ are {\it iid}\/ uniform among $\{\frac{\pm1\pm j}{\sqrt{2}}\}$.
\begin{equation*}
    \cG_{3,\rm O}(n,\hat{r})=-2\ln\E_{\bm}\exp(\sum_{a<b}\hat{r}m_am_b),
\end{equation*}
where $\bm=[m_0,\cdots,m_n]^T$ with each element {\it iid}\/ uniform in $\{\pm1\}$. According to \cite{tanaka2002statistical}, we have
\begin{equation}
    \cG_{3,\rm O}(n,\hat{r})=n\hat{r}-2\ln\E_u\cosh^n(\hat{r}+\sqrt{\hat{r}}u),
\end{equation}
where $u\sim\cN(0,1)$.

\item Derive $\cF_2(r,\hat{r})$ according to (\ref{eq:F2_def}):

For linear transmitters, we have
\begin{align*}
    \cF_{2,\rm L}(r,\hat{r})=&-4\ratio\E_{u}Q(A\sqrt{r}u)\ln Q(A\sqrt{r}u)\\
    &+\ln(1+\hat{r})-\hat{r}+r\hat{r},
\end{align*}
where 
\begin{equation}
    A=\sqrt{\frac{\snreff }{\snreff(1-r)+1}}.
\end{equation}

For one-bit transmitters, we have
\begin{align*}
    \cF_{2,\rm O}(r,\hat{r})&=-4\ratio\E_{u}Q(A\sqrt{r}u)\ln Q(A\sqrt{r}u)+\hat{r} \\
    &-2\E_u\ln \cosh(\hat{r}+\sqrt{\hat{r}}u)+r\hat{r}.
\end{align*}

\item Solve for ($\qx,\qxhat$) according to (\ref{eq:qx_RS_solution}):

For linear transmitters, we get that ($\qxL,\qxLhat$) are the solution of 
\begin{equation}
\qtxhat = \frac{\ratio A^2}{\pi}\E_u
\frac{\exp\left(-A^2\qtx u^2\right)}{Q(A\sqrt{\qtx}u)},
\label{eq:q_hat_solution}
\end{equation}
\begin{equation}
\qtx = \frac{\qtxhat}{1+\qtxhat},A = \sqrt{\frac{\snreff}{1+\snreff(1-\qtx)}}.
\label{eq:A_solution}
\end{equation} 

For one-bit transmitters, we get that
($\qxO,\qxOhat$) are the solution of 
\begin{equation}
\qxhat = \frac{\ratio A^2}{\pi}\E_u \frac{\exp\left(-A^2\qx u^2\right)}{Q(A\sqrt{\qx}u)},
\label{eq:qx_hat_solution_one_bit_tx_rx_RS}
\end{equation}
\begin{equation}
\qx+1 = \E_u\tanh(\sqrt{\qxhat}u+\qxhat)(2+\frac{u}{\sqrt{\qxhat}}),
\label{eq:qx_solution_one_bit_tx_rx_RS}
\end{equation}
\begin{equation*}
    A = \sqrt{\frac{\snreff}{1+\snreff(1-\qx)}}.
\end{equation*}

\item Compute $\Ravgeff(\beta_{\rm t,k})$ according to (\ref{eq:Reff_RS_solution}):

For linear transmitters, we have
\begin{align*}
    &\RavgeffL(\beta_{\rm t,k}) = \frac{1}{\ln2}\Big[\cF_{2,\rm L}(\qxL,\qxLhat) \\
    &+ 4\ratio\E_uQ(\sqrt{\snreff}u)\ln Q(\sqrt{\snreff}u)\Big],u\sim\cN(0,1).
\end{align*}

For one-bit transmitters, we have
\begin{align*}
    &\RavgeffO(\beta_{\rm t,k}) = \frac{1}{\ln2}\Big[\cF_{2,\rm O}(\qxO,\qxOhat) \\
    &+ 4\ratio\E_uQ(\sqrt{\snreff}u)\cdot\ln Q(\sqrt{\snreff}u)\Big],u\sim\cN(0,1).
    \numberthis
    \label{eq:Reff_linear_tx_RS}
\end{align*}
This result matches that in \cite{gao2018capacity}. 

\item Compute $\Cavg_{\rm bound}$ and $\betatopt$

For linear transmitters, we have
\begin{equation}
    \betatoptL=\argmax_{\beta_{{\rm t},k}}\frac{\beta-\beta_{{\rm t},k}}{\beta}\RavgeffL(\beta_{{\rm t},k}).
    \label{eq:betat_opt_L_RS}
\end{equation}
\begin{equation}
    \Cavg_{\rm{bound,L}}=\frac{\beta-\betatoptL}{\beta}\RavgeffL(\betatoptL).
    \label{eq:lower_bound_L_RS}
\end{equation}

For one-bit transmitters, we have
\begin{equation}
    \betatoptO=\argmax_{\beta_{{\rm t},k}}\frac{\beta-\beta_{{\rm t},k}}{\beta}\RavgeffO(\beta_{{\rm t},k}).
    \label{eq:betat_opt_O_RS}
\end{equation}
\begin{equation}
    \Cavg_{\rm{bound,O}}=\frac{\beta-\betatoptO}{\beta}\RavgeffO(\betatoptO),
    \label{eq:lower_bound_one_bit_RS}
\end{equation}

\end{enumerate}

\section{Numerical Results}
We provide some numerical results based on the expressions we obtained for two types of nonlinear systems in previous section. We select an accuracy of 0.1 in our optimization over $\betat$.

\subsection{Linear transmitters and one-bit receivers}
For a system with linear transmitters and one-bit receivers, a training-based capacity lower bound using the Bussgang theorem is derived in \cite{li2016much,li2017channel}. LMMSE channel estimation is used and the channel estimation error part is approximated at low SNR to obtain a closed-form solution. However, with our equivalent channel (\ref{eq:equivalent_noise_model}), the Bussgang decomposition can be applied to get a simplified solution for any SNR. In our equivalent channel model, MMSE channel estimation is used, which provides better performance than LMMSE channel estimation in general. The lower bound thereby obtained is
\begin{equation}
    \Cavg_{\rm t}\geq\CavgBus =\max_{\betat}\frac{\beta-\betat}{\beta} \log_2\left(1+\frac{2\ratio\snreff}{\pi(1+\snreff)}\right),
    \label{eq:bussgang_rate}
\end{equation}
where $\snreff$ is defined in (\ref{eq:effective_SNR}). 

For comparison, we also consider the achievable rate when the receiver knows the channel and each element of $\txvd$ are {\it iid}\/ $\cC\cN(0,1)$, denoted as $\RavgCSIR$. $\RavgCSIR$ can be directly derived using steps from 5B) to 9B) with $\qh=1$. And the solution of $\RavgCSIR$ is
\begin{align*}
    &\RavgCSIR=\frac{1}{\ln2}\big[4\ratio\E_{u}\big(Q(\sqrt{\snr}u)\ln Q(\sqrt{\snr}u) \\
    &- Q(\hat{A}\sqrt{\qx}u)\ln Q(\hat{A}\sqrt{\qx}u)\big)+\ln(1+\qxhat)-\qxhat+\qx\qxhat\big],
    \numberthis
    \label{eq:Ravg_CSIR}
\end{align*}
where $\hat{A},\qx,\qxhat$ are the solution of
\begin{equation}
\qtxhat = \frac{\ratio \hat{A}^2}{\pi}\E_u
\frac{\exp\left(-\hat{A}^2\qtx u^2\right)}{Q(\hat{A}\sqrt{\qtx}u)},
\end{equation}
\begin{equation}
\qtx = \frac{\qtxhat}{1+\qtxhat},\hat{A} = \sqrt{\frac{\snr}{1+\snr(1-\qtx)}}.
\end{equation} 

\begin{figure}
\includegraphics[width=3.5in]{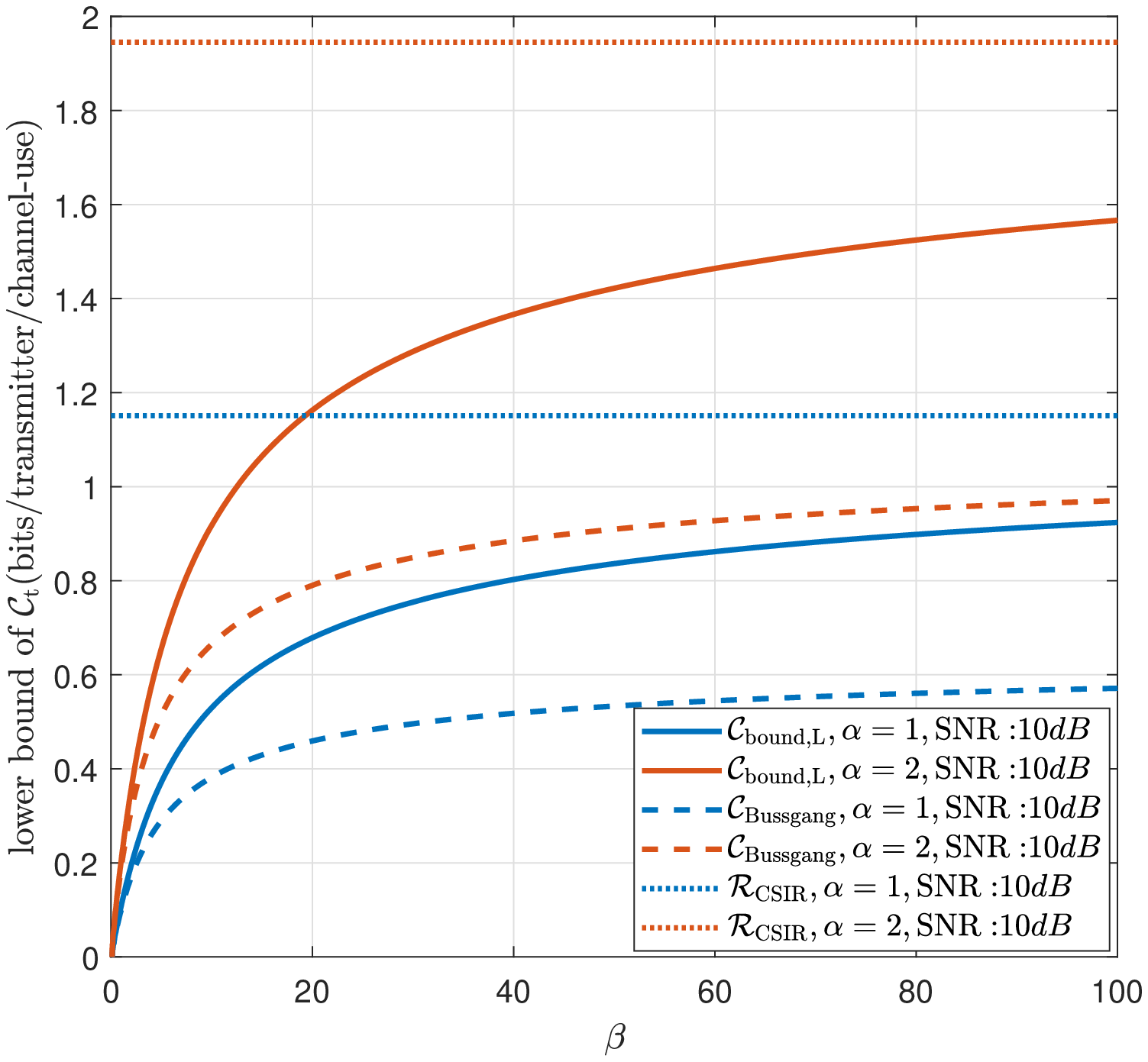}
\centering
    \caption{The comparison between $\Cavg_{\rm bound,L}$ (\ref{eq:lower_bound_L_RS}) and $\CavgBus$ (\ref{eq:bussgang_rate})  at $\ratio=1,2$ with different $\beta$ for systems with linear transmitters and one-bit receivers. Both $\CavgBus$ and $\Cavg_{\rm bound,L}$ are maximized over $\betat$. $\RavgCSIR$ (\ref{eq:Ravg_CSIR}) is the achievable rate when the receiver knows the channel and the elements in $\txvd$ are {\it iid}\/ $\cC\cN(0,1)$.}
    \label{fig:RS_vs_Bussgang_CSIR}
\end{figure}

The comparison between $\CavgBus,\Cavg_{\rm bound,L}$ and $\RavgCSIR$ is shown in Fig. \ref{fig:RS_vs_Bussgang_CSIR}. We observe that $\CavgBus$ is generally smaller (less tight as a lower bound) than $\Cavg_{\rm bound,L}$.

When $\snreff\to 0$, which can be caused by small $\betat$ or small $\snr$, $\CavgBus$ and $\Cavg_{\rm bound,L}$ meet.  From (\ref{eq:q_hat_solution}) and (\ref{eq:A_solution}), we have
\begin{equation}
    \qxL=\frac{2\ratio}{\pi}\snreff+o(\snreff),\qxLhat=\frac{2\ratio}{\pi}\snreff+o(\snreff).
\end{equation}
Then, (\ref{eq:lower_bound_L_RS}) becomes 
\begin{equation}
    \Cavg_{\rm bound,L}=\max_{\betat}\frac{\beta-\betat}{\beta}\frac{2\alpha}{\pi\ln2}\snreff+o(\snreff),
\end{equation}
which is the same as $\CavgBus$.  According to (\ref{eq:qh_solution_one_bit_rx}), for small SNR with $\snr\to 0$, we have
\begin{equation}
\qh=\frac{2\betat\snr}{\pi}+o(\snr).
\end{equation}
Therefore, (\ref{eq:effective_SNR}) becomes
\begin{equation}
    \snreff=\frac{2\betat\snr^2}{\pi}+o(\snr^2).
    \label{eq:snreff_small}
\end{equation}
And we have
\begin{equation*}
    \betatoptL \approx \frac{\beta}{2},\quad\Cavg_{\rm{bound},L} \approx \frac{\ratio\beta}{\pi^2\ln2}\snr^2,
\end{equation*}
when $\snr\to 0$.  Hence, half of $\Tb$ is used for training at low SNR with linear transmitters and one-bit receivers.

\subsection{One-bit transmitters and one-bit receivers}
\begin{figure}
\includegraphics[width=3.5in]{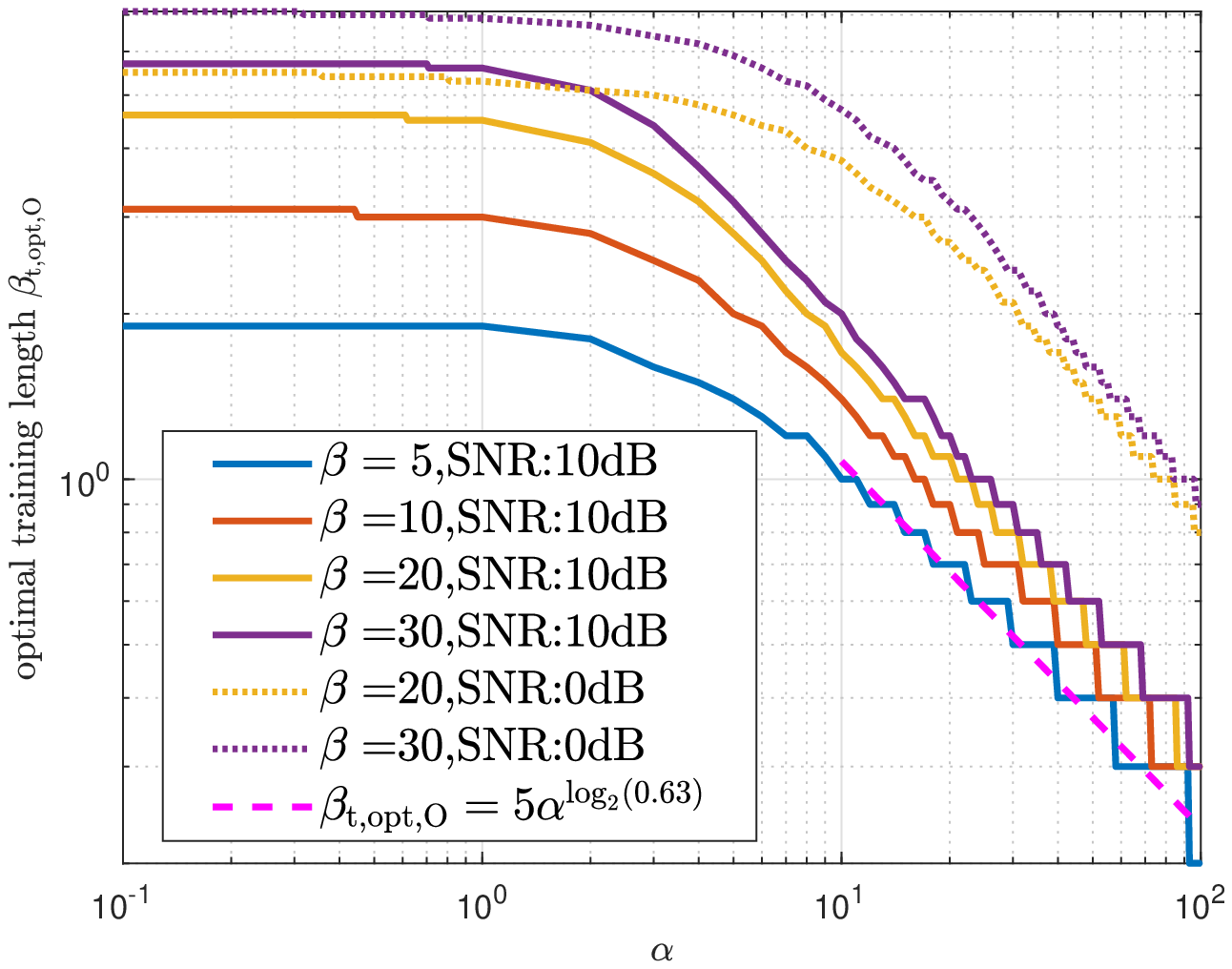}
\centering
    \caption{The optimal training time  $\betatoptO$ (\ref{eq:betat_opt_O_RS}) versus $\ratio=\frac{\rn}{\tn}$ in a wide range. We can see that it's possible to have $\betatoptO<1$. Both $\ratio$ and $\betatoptO$ are plotted in a log domain. When $\alpha$  is large, $\betatoptO$ decreases by 37 percent when we double $\alpha$.}
    \label{fig:opt_beta_t_SNR10_0}
\end{figure}

For systems with one-bit transceivers, we show that it is possible to have optimal training time smaller than the number of transmitter ($\betatopt<1$) even when the total time is larger than twice of the number of transmitters ($\beta>2$). This result is shown in Fig. \ref{fig:opt_beta_t_SNR10_0}, where we show the relationship between $\betatoptO$ and $\ratio$ with different $\beta$. We see that when $\ratio$ is large, $\betatoptO$ can be smaller than 1 and $\betatoptO$ decreases approximately 37 percent when we double $\alpha$ as $\alpha$ gets large. When $\ratio$ is small, according to \cite{gao2018capacity}, 
\begin{equation}
    \RavgeffO \approx \ratio c(\snreff),
\end{equation}
where $c(\snr)$ is the capacity of a single pair of one-bit transceivers in a Rayleigh channel with channel information at the receiver at SNR $\snr$, defined as
\begin{equation}
c(\snr) = 2(1-\E_z\left(\Entfuntwo(Q(\sqrt{\snr} z))\right)), z\sim\cN(0,1), 
\end{equation}
where $\Entfuntwo(p)=-(p\log_2p + (1-p)\log_2(1-p))$ is the binary entropy function. 

\begin{figure}
\includegraphics[width=3.5in]{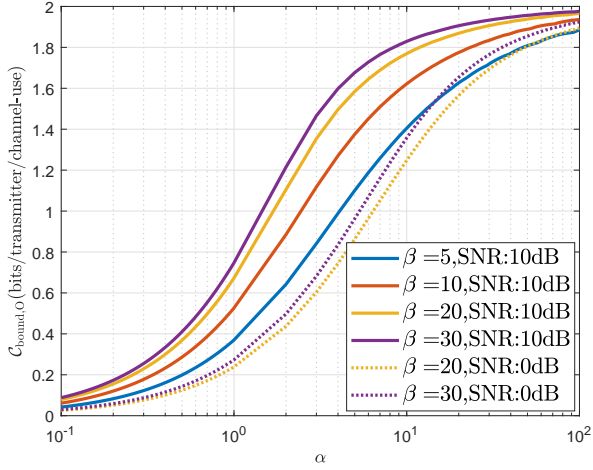}
\centering
    \caption{The lower bound $\Cavg_{\rm bound,O}$ (\ref{eq:lower_bound_one_bit_RS}) versus  $\ratio=\frac{\rn}{\tn}$ in a wide range. Because of the one-bit quantization at the transmitter, the rate per transmitter can not be more than 2. Therefore, we can clearly see the saturation effect when we increase $\ratio$.}
    \label{fig:low_bound_vs_alpha}
\end{figure}

The $\betatoptO$ obtained for small $\ratio$ maximize $\frac{\beta-\betat}{\beta}c(\snreff)$, where $\snreff$ is solved from (\ref{eq:effective_SNR}).
The corresponding bound $\Cavg_{\rm bound,O}$ is shown in Fig. \ref{fig:low_bound_vs_alpha}. Because of the one-bit quantization at the transmitter, $\Cavg_{\rm bound,O}$ saturates at $\Cavg_{\rm bound,O}=2$ as $\ratio$ gets larger. With many more receivers than the transmitters ($\alpha$ very large), saturation is achieved with a small number of training symbols, and hence $\betatoptO$ can be smaller than 1.

At low SNR, $\snreff\to 0$, according to (\ref{eq:qx_hat_solution_one_bit_tx_rx_RS}) and (\ref{eq:qx_solution_one_bit_tx_rx_RS}), we have
\begin{equation}
    \qxO=\frac{2\ratio}{\pi}\snreff+o(\snreff),\qxOhat=\frac{2\ratio}{\pi}\snreff+o(\snreff),
\end{equation}
and (\ref{eq:lower_bound_one_bit_RS}) becomes
\begin{equation}
    \Cavg_{\rm bound,O}=\max_{\betat}\frac{\beta-\betat}{\beta}\frac{2\alpha}{\pi\ln2}\snreff+o(\snreff).
\end{equation}
According to (\ref{eq:snreff_small}), we have
\begin{equation*}
    \betatoptO \approx \frac{\beta}{2},\quad\Cavg_{\rm{bound},O} \approx \frac{\ratio\beta}{\pi^2\ln2}\snr^2.
\end{equation*}
Thus, as with linear transmitters, half of $\Tb$ is used for training at low SNR with one-bit transmitters and receivers.

\appendices
\section{mutual information equivalence}
\label{app:mutual_information_equivalence}
We show the main steps to prove our claim using replica method. Some techniques we use are similar to those used in \cite{wen2016bayes}.
According to (\ref{eq:Reff_RS_def}), we have
\begin{equation}
    \Ravgeff=\lim_{\tn\to\infty}\frac{1}{\tn}\left(\Ent(\rxvd|\txmt,\rxmt)-\Ent(\rxvd|\txmt,\rxmt,\txvd)\right)
\end{equation}
Similar to \cite{wen2016bayes}, we apply "replica trick" and have
\begin{align*}
    &\lim_{\tn\to\infty}\frac{1}{\tn}\Ent(\rxvd|\txmt,\rxmt)=\lim_{\tn\to\infty}\frac{-1}{\tn\ln2}\E\ln\frac{p(\rxvd,\rxmt|\txmt)}{p(\rxmt|\txmt)}\\
    &=-\lim_{\tn\to\infty}\frac{1}{\tn\ln2}\lim_{n\to 0}\frac{\partial}{\partial n}\left[\ln\Xi_n-\ln\hat{\Xi}_n\right],
\end{align*}
where
\begin{equation*}
    \Xi_n = \E[p^n(\rxvd,\rxmt|\txmt)],\hat{\Xi}_n = \E[p^n(\rxmt|\txmt)].
\end{equation*}

We assume the limit of $n$ and $\tn$ can commute and we have
\begin{align*}
    &\lim_{\tn\to\infty}\frac{1}{\tn\ln2}\Ent(\rxvd|\txmt,\rxmt)=\\
    &-\lim_{n\to 0}\frac{\partial}{\partial n}\lim_{\tn\to\infty}\frac{1}{\tn\ln2}\left[\ln\Xi_n-\ln\hat{\Xi}_n\right].
    \numberthis
    \label{eq:replica_Hy_training}
\end{align*}

Also, we consider $n$ as integer to derive $\Xi_n$ and $\hat{\Xi}_n$ as a function of $n$, and we assume the expression still holds for real number $n$. Then, for integer $n$, we have
\begin{align*}
    &\Xi_n = \E[p^n(\rxvd,\rxmt|\txmt)]\\
    &=\E_{\txmt}\sum_{\rxmt,\rxvd}\left(\E_{\txvd,\chm}p(\rxvd,\rxmt|\txvd,\txmt,\chm)\right)^{n+1}\\
    &=\E_{\txmt,[\chm]_{0}^{n},[\txvd]_{0}^{n}}\prod_{k=1}^{\rn}\sum_{\rx_{\text{d},k}}\prod_{a=0}^n\gfunc\left(\sqrt{\frac{\snr}{\tn}}\chv^{(a)T}_k\txvd^{(a)},\rx_{\text{d},k},\sigma^2\right)\\
    &\cdot \prod_{k=1}^{\rn}\prod_{p=1}^{\Tt}\sum_{\rx_{\text{t},kp}}\prod_{a=0}^{n}\gfunc\left(\sqrt{\frac{\snr}{\tn}}\chv^{(a)T}_k\txvta[p],\rx_{\text{t},kp},\sigma^2\right)\\
    &=\E_{\txmt,[\txvd]_{0}^{n}}\Bigg[\E_{[\chv]_{0}^{n}}\Big[\sum_{\rx_{\text{d}}}\prod_{a=0}^n\gfunc\left(\sqrt{\frac{\snr}{\tn}}\chv^{(a)T}\txvd^{(a)},\rx_{\text{d}},\sigma^2\right) \\
    & \cdot\prod_{p=1}^{\Tt}\sum_{\rx_{\text{t},p}}\prod_{a=0}^{n}\gfunc\left(\sqrt{\frac{\snr}{\tn}}\chv^{(a)T}\txvta[p],\rx_{\text{t},p},\sigma^2\right)\Big] \Bigg]^\rn
\end{align*}
where $\rx_{\text{t},kp}$ is the $k$th row and $p$th column of $\rxmt$, $\txv_{\text{t},p}$ is the $p$th column of $\txmt$, $\chv_k^{(a)T}$ is the $k$th row of $\chm^{(a)}$,  $[\chm]_{0}^{n}=\{\chm^{(0)},\cdots,\chm^{(n)}\}$,$[\chv]_{0}^{n}=\{\chv^{(0)},\cdots,\chv^{(n)}\}$,$[\txvd]_{0}^{n}=\{\txvd^{(0)},\cdots,\txvd^{(n)}\}$, which are collection of replicas. We drop $k$ at the last step because each element in $\chm$ are {\it iid}.

We introduce two $(n+1)\times(n+1)$ matrices $Q_{\rm \chs}=[q_{\rm \chs}^{ab}]$ and $Q_{\rm \tx}=[q_{\rm \tx}^{ab}]$ whose elements are defined as $q_{\rm \chs}^{ab}=\frac{(\chv^{(a)})^H\chv^{(b)}}{\tn}$, $q_{\rm \tx}^{ab} = \frac{(\txvd^{(a)})^H\txvd^{(b)}}{\tn}$. Let $z^{(a)}_p=\sqrt{\frac{1}{\tn}}\chv^{(a)T}\txvta[p]$, $\bz_p=[z_p^{(0)},\cdots,z_p^{(n)}]^T$, $z_{\rm{d}}^{(a)}=\sqrt{\frac{1}{\tn}}\chv^{(a)T}\txvd^{(a)}$, $\bz_{\rm{d}}=[z_{\rm{d}}^{(0)},\cdots,z_{\rm{d}}^{(n)}]^T$, then $\bz_p\sim\cC\cN(0,Q_{\rm \chs})$, $\bz_{\rm{d}}\sim\cC\cN(0,Q_z)$ for large $\tn$ where $Q_z=(Q_{\rm \chs}\circ Q_{\rm \tx})$ is the Hadamard product between $Q_{\rm \chs}$ and $Q_{\rm \tx}$. Since $\txvd^{(a)}$ is independent of $\txmt$, and elements of $\txmt$ are {\it iid}\/, we have
\begin{align*}
    &\Xi_n = \E_{[\txvd]_{0}^{n}}\Bigg[\E_{[\chv]_{0}^{n}}\Big[\E_{{\bz_{\rm{d}}}}\sum_{\rx_{\text{d}}}\prod_{a=0}^n\gfunc\left(\sqrt{\snr}z_{\rm{d}}^{(a)},\rx_{\text{d}},\sigma^2\right) \\
    & \cdot\prod_{p=1}^{\Tt}\E_{\bz_p}\sum_{\rx_{\rm {t}, p}}\prod_{a=0}^{n}\gfunc\left(\sqrt{\snr}z^{(a)}_p,\rx_{\rm{t},p},\sigma^2\right) \Big]\Bigg]^\rn\\
    &= \E_{[\txvd]_{0}^{n}}\Bigg[\E_{[\chv]_{0}^{n}}\Big[\E_{{\bz_{\rm{d}}}}\sum_{\rx_{\text{d}}}\prod_{a=0}^n\gfunc\left(\sqrt{\snr}z_{\rm{d}}^{(a)},\rx_{\text{d}},\sigma^2\right) \\
    & \cdot\Big[\E_{\bz}\sum_{\rxt}\prod_{a=0}^{n}\gfunc\left(\sqrt{\snr}z^{(a)},\rxt,\sigma^2\right) \Big]^{\Tt}\Big]\Bigg]^{\rn},
\end{align*}
where $\bz\sim\cC\cN(0,Q_{\rm \chs})$.

Let 
\begin{align*}
    \mathcal{J}_1(Q_{\rm \chs})&=\E_{\bz}\sum_{\rxt}\prod_{a=0}^{n}\gfunc\left(\sqrt{\snr}z^{(a)},\rxt,\sigma^2\right),\\
    \numberthis
    \label{eq:J1_Qh}
    \mathcal{J}_2(Q_{\rm \chs},Q_{\rm \tx})&=\E_{{\bz_{\rm{d}}}}\sum_{\rx_{\text{d}}}\prod_{a=0}^n\gfunc\left(\sqrt{\snr}z_{\rm{d}}^{(a)},\rx_{\text{d}},\sigma^2\right).
\end{align*}
Then
\begin{equation}
    \Xi_n = \E_{[\txvd]_{0}^{n}}\Bigg[\E_{[\chv]_{0}^{n}}\Big[[\mathcal{J}_1(Q_{\rm \chs})]^{\Tt}\mathcal{J}_2(Q_{\rm \chs},Q_{\rm \tx})\Big]\Bigg]^{\rn}.
\end{equation}

Similarly, we have
\begin{align*}
    &\hat{\Xi}_n = \E[p^n(\rxmt|\txmt)]\\
    &=\E_{\txmt}\Bigg[\E_{[\chv]_{0}^{n}} \prod_{p=1}^{\Tt}\sum_{\rx_{\text{t},p}}\prod_{a=0}^{n}\gfunc\left(\sqrt{\frac{\snr}{\tn}}\chv^{(a)T}\txvta[p],\rx_{\text{t},p},\sigma^2\right)\Bigg]^\rn\\
    &=\Bigg[\E_{[\chv]_{0}^{n}}\Big[\E_{\bz}\sum_{\rxt}\prod_{a=0}^{n}\gfunc\left(\sqrt{\snr}z^{(a)},\rxt,\sigma^2\right) \Big]^{\Tt}\Bigg]^{\rn}\\
    &=\Bigg[\E_{[\chv]_{0}^{n}}[\mathcal{J}_1(Q_{\rm \chs})]^{\Tt}\Bigg]^{\rn}.
\end{align*}

Therefore,
\begin{equation*}
 \ln\Xi_n-\ln\hat{\Xi}_n=\ln\E_{[\txvd]_{0}^{n}}\Bigg[\frac{\E_{[\chv]_{0}^{n}}\Big[[\mathcal{J}_1(Q_{\rm \chs})]^{\Tt}\mathcal{J}_2(Q_{\rm \chs},Q_{\rm \tx})\Big]}{\E_{[\chv]_{0}^{n}}[\mathcal{J}_1(Q_{\rm \chs})]^{\Tt}}\Bigg]^{\rn}
\end{equation*}

When $\Tt\to\infty$, based on the saddle point method,
\begin{equation}
\frac{\E_{[\chv]_{0}^{n}}\Big[[\mathcal{J}_1(Q_{\rm \chs})]^{\Tt}\mathcal{J}_2(Q_{\rm \chs},Q_{\rm \tx})\Big]}{\E_{[\chv]_{0}^{n}}[\mathcal{J}_1(Q_{\rm \chs})]^{\Tt}}\to \mathcal{J}_2(\tilde{Q}_{\rm \chs},Q_{\rm \tx}),
    \label{eq:saddle_point_qh}
\end{equation}
where $\tilde{Q}_{\rm \chs}$ is the saddle point of $\E_{[\chv]_{0}^{n}}[\mathcal{J}_1(Q_{\rm \chs})]^{\Tt}$. We still need to keep it in mind that the saddle point should be considered at the derivative of $n$ with $n\to 0$. Therefore, $\tilde{Q}_{\rm \chs}$ can be obtained by solving
\begin{equation}
    \cF_{\rm \chs}=-\lim_{n\to 0}\frac{\partial}{\partial n}\lim_{\tn\to\infty}\frac{1}{\tn}\ln \E_{[\chv]_{0}^{n}}[\mathcal{J}_1(Q_{\rm \chs})]^{\Tt}.
    \label{eq:F_h_RS}
\end{equation}

It is not hard to show 
\begin{equation}
    -\lim_{\tn\to\infty}\frac{1}{\tn\rn}\E_{\txmt,\rxmt}\ln p(\rxmt|\txmt)=\cF_{\rm \chs}
    \label{eq:training_RS_saddle_point}
\end{equation}
through regular steps used in replica method with similar assumptions.

Now, we use replica symmetry (RS) assumption by assuming the off-diagonal elements of the saddle point $\tilde{Q}_{\rm \chs}$ are equal, denoted as $\qh$. The diagonal elements of $\tilde{Q}_{\rm \chs}$ are 1, which is the variance of the elements of channel. According to \cite{opper1996statistical,wen2016bayes}, when we obtain the saddle point $\qh$ through (\ref{eq:training_RS_saddle_point}), $1-\qh$ describes the MSE of the MMSE channel estimation, shown in (\ref{eq:channel_MMSE_RS}). 

Then, we have
\begin{align*}
    &\lim_{\tn\to\infty}\frac{1}{\tn}\left[\ln\Xi_n-\ln\hat{\Xi}_n\right]\\
    &=\lim_{\tn\to\infty}\frac{1}{\tn}\ln\E_{[\txvd]_{0}^{n}}\left(\mathcal{J}_2(\tilde{Q}_{\rm \chs},Q_{\rm \tx})\right)^\rn\\
    &=\lim_{\tn\to\infty}\frac{1}{\tn}\ln\E_{[\txvd]_{0}^{n}}\Bigg[\sum_{\rxd}\E_{\bz_{\rm{d}}}\prod_{a=0}^{n}\gfunc(\sqrt{\snr}z_{\rm{d}}^{(a)},\rxd,\sigma^2)\Bigg]^\rn,
\end{align*}
where $\bz_{\rm{d}}\sim\cC\cN(0,Q_z)$, $Q_z=\tilde{Q}_{\rm \chs}\circ Q_{\rm \tx}$.

Therefore,
\begin{align*}
    &\lim_{\tn\to\infty}\frac{1}{\tn}\Ent(\rxvd|\txmt,\rxmt)=-\lim_{n\to 0}\frac{\partial}{\partial n}\lim_{\tn\to\infty}\frac{1}{\tn\ln2}\\
    &\ln\E_{[\txvd]_{0}^{n}}\Bigg[\sum_{\rxd}\E_{\bz_{\rm{d}}}\prod_{a=0}^{n}\gfunc(\sqrt{\snr}z_{\rm{d}}^{(a)},\rxd,\sigma^2)\Bigg]^\rn.
\end{align*}

Similarly, we have
\begin{align*}
    &\lim_{\tn\to\infty}\frac{1}{\tn}\Ent(\rxvd|\txmt,\rxmt,\txvd)=-\lim_{n\to 0}\frac{\partial}{\partial n}\lim_{\tn\to\infty}\frac{1}{\tn\ln2}\\
    &\ln\Bigg[\sum_{\rxd}\E_{\bz_\chs}\prod_{a=0}^{n}\gfunc(\sqrt{\snr}z_\chs^{(a)},\rxd,\sigma^2)\Bigg]^\rn,
\end{align*}
where $\bz_\chs=[z_\chs^{(0)},\cdots,z_\chs^{(n)}]^T$ and $\bz_\chs\sim\cC\cN(0,\tilde{Q}_\chv)$.

For the equivalent channel with known $\hat{\chm}$ shown in (\ref{eq:equivalent_noise_model}), similarly, we have  \begin{align*}
    &\lim_{\tn\to\infty}\frac{1}{\tn}\Ent(\rxvdhat|\hat{\chm})=-\lim_{n\to 0}\frac{\partial}{\partial n}\lim_{\tn\to\infty}\frac{1}{\tn\ln2}\\
    &\ln\E_{[\txvd]_{0}^{n}}\Bigg[\sum_{\rxd}\E_{\hat{\bz}_{\rm d}}\prod_{a=0}^{n}\gfunc(\sqrt{\snr}\hat{z}_{\rm d}^{(a)},\rxd,\hat{\sigma}^2)\Bigg]^\rn,
    \numberthis
    \label{eq:entropy_eff_H}
\end{align*}
where $\hat{\bz}_{\rm d}=[\hat{z}_{\rm d}^{(0)},\cdots,\hat{z}_{\rm d}^{(n)}]^T\sim\cC\cN(0,\qh\cdot Q_{\rm \tx})$, $\hat{\sigma}^2=\sigma^2+\snr(1-\qh)$. The joint distribution of $\sqrt{\snr}\hat{\bz}_{\rm d}+\hat{\nv}$ is the same as the joint distribution of $\sqrt{\snr}\bz_{\rm{d}}+\nv$. According to (\ref{eq:gfunc_discrete_out}) and (\ref{eq:gfunc_continuous_out}), we have
\begin{equation*}
    \E_{\bz_{\rm{d}}}\prod_{a=0}^{n}\gfunc(\sqrt{\snr}z_{\rm{d}}^{(a)},\rxd,\sigma^2) = \E_{\hat{\bz}_{\rm d}}\prod_{a=0}^{n}\gfunc(\sqrt{\snr}\hat{z}_{\rm d}^{(a)},\rxd,\hat{\sigma}^2),
\end{equation*}
and therefore
\begin{equation*}
    \lim_{\tn\to\infty}\frac{1}{\tn}\Ent(\rxvd|\txmt,\rxmt) = \lim_{\tn\to\infty}\frac{1}{\tn}\Ent(\rxvdhat|\hat{\chm}).
\end{equation*}

Similarly, we have
\begin{equation*}
    \lim_{\tn\to\infty}\frac{1}{\tn}\Ent(\rxvd|\txmt,\rxmt,\txvd) = \lim_{\tn\to\infty}\frac{1}{\tn}\Ent(\rxvdhat|\hat{\chm},\txvd).
\end{equation*}
Therefore,
\begin{equation*}
 \lim_{\tn\to\infty}\frac{1}{\tn}\MuI(\txvd;\rxvd|\txmt,\rxmt)=\lim_{\tn\to\infty}\frac{1}{\tn}\MuI(\txvd;\rxvdhat|\hat{\chm}),
\end{equation*}
where $\txvd$ and $\rxvdhat$ are the input and output of channel defined in (\ref{eq:equivalent_noise_model}).

\section{proof of replica method to compute $\Cavg_{\rm{bound}}$}
\label{app:channel_estimation_MI_RS}
According to Appendix \ref{app:mutual_information_equivalence}, $\qh$ can be obtained by solving $\cF_{\rm \chs}$ defined in (\ref{eq:F_h_RS}).

Similarly to \cite{tanaka2004statistical}, we apply Varadhan's theorem and Gartner-Ellis theorem\cite{dembo38large} and obtain
\begin{align*}
    &\lim_{\tn\to\infty}\frac{1}{\tn}\ln \E_{[\chv]_{0}^{n}}[\mathcal{J}_1(Q_{\rm \chs})]^{\Tt} = \sup_{Q_{\rm \chs}}\inf_{\hat{Q}_{\rm \chs}}\Big[
\betat G_1(Q_{\rm \chs}) \\
&- \sum_{a<b}2q_{\rm \chs}^{ab}\hat{q}_{\rm \chs}^{ab} + L_1(\hat{Q}_{\rm \chs})\Big],
\end{align*}
where $q_{\rm \chs}^{ab}$ and $\hat{q}_{\rm \chs}^{ab}$ are elements of $Q_{\rm \chs}$ and $\hat{Q}_{\rm \chs}$,
\begin{equation}
    G_1(Q_{\rm \chs})=\ln\left(\mathcal{J}_1(Q_{\rm \chs})\right),
\end{equation}
\begin{equation*}
    L_1(\hat{Q}_{\rm \chs})=\lim_{\tn\to\infty}\frac{1}{\tn}\ln\E_{[\chv]_0^{n}}\exp\left(\sum_{a<b}2\qhhat^{ab}(\chv^{(a)})^H\chv^{(b)}\right).
\end{equation*}

The values of $Q_{\rm \chs}$ and $\hat{Q}_{\rm \chs}$ that achieves the extremum are called saddle point. Based on the RS assumption, we assume the off-diagonal elements of $Q_{\rm \chs}$ and $\hat{Q}_{\rm \chs}$ are the same, denoted as $\qh$ and $\qhhat$, respectively. Then, we have
\begin{equation*}
    \betat G_1(Q_{\rm \chs})=-\cG_1(n,\qh),
\end{equation*}
where $\cG_1(n,q)$ is defined in (\ref{eq:G1_def}). Also,
\begin{equation*}
L_1(\hat{Q}_{\rm \chs}) =  \mathcal{L}_1(n,\qhhat),
\end{equation*}
where
\begin{equation*}
    \mathcal{L}_1(n,\qhhat) = -(n+1)\ln(1+\qhhat) - \ln\left(1-\frac{\qhhat(n+1)}{1+\qhhat}\right).
\end{equation*}
Therefore, (\ref{eq:F_h_RS}) becomes
\begin{equation*}
    \cF_{\rm \chs} = \cF_1(\qh,\qhhat),
\end{equation*}
where $\cF_1(q,\hat{q})$ is defined in (\ref{eq:cF1_value}). $(\qh,\qhhat)$ is the saddle point of $\cF_1(q,\hat{q})$ and we have
\begin{equation}
    \frac{\partial \cF_1(q,\hat{q})}{\partial q}=0, \frac{\partial \cF_1(q,\hat{q})}{\partial \hat{q}}=0,
\end{equation}
at $(\qh,\qhhat)$. If there are multiple solutions, we should use the one that minimize $\cF_1(\qh,\qhhat)$. We finish the proof of solving $\qh$ in steps 2B)-4B) shown in the recipe. $\qh$ depends on $\betat$, and for the rest of the steps, only $\qh$ is needed to compute $\Ravgeff(\betat)$.

We use the equivalent channel shown in (\ref{eq:Reff_equivalence}) to compute $\Ravgeff(\betat)$ and we have
\begin{equation}
    \Ravgeff(\betat)=\lim_{\tn\to\infty}\frac{1}{\tn}(\Ent(\rxvdhat|\hat{\chm}) - \Ent(\rxvdhat|\hat{\chm},\txvd)).
\end{equation}

Since $\sqrt{\frac{\snr}{\tn}}\hat{\chm}\txvd\sim\cC\cN(0,\snr\qh\cdot I)$, we have
\begin{align*}
    \lim_{\tn\to\infty}&\frac{1}{\tn}\Ent(\rxvdhat|\hat{\chm},\txvd))=\frac{\ratio}{\ln2}\sum_{\rx\in\rxA}\E_{z}\Big(
    \gfunc(\sqrt{\snr\qh}z,\rx,\hat{\sigma}^2)\\
    &\cdot \ln\gfunc(\sqrt{\snr\qh}z,\rx,\hat{\sigma}^2)\Big), z\sim\cC\cN(0,1).
\end{align*}

According to (\ref{eq:entropy_eff_H}), we have
\begin{equation}
    \lim_{\tn\to\infty}\frac{1}{\tn}\Ent(\rxvdhat|\hat{\chm}) = -\lim_{n\to 0}\frac{\partial}{\partial n}\lim_{\tn\to\infty}\frac{1}{\tn\ln2}\ln\Xi_{\rm{\tx},n},
    \label{eq:Entropy_RS_effective}
\end{equation}
where
\begin{align*}
    \Xi_{\rm{\tx},n}&=\E_{[\txvd]_{0}^{n}}\Bigg[\sum_{\rxd}\E_{\hat{\bz}_{\rm d}}\prod_{a=0}^{n}\gfunc(\sqrt{\snr}\hat{z}_{\rm d}^{(a)},\rxd,\hat{\sigma}^2)\Bigg]^\rn\\
    &=\E_{[\txvd]_{0}^{n}}\Bigg[\sum_{\rxd}\E_{\bw}\prod_{a=0}^{n}\gfunc(\sqrt{\snr\qh}w^{(a)},\rxd,\hat{\sigma}^2)\Bigg]^\rn,
\end{align*}
with $\bw=[w^{(0)},\cdots,w^{(n)}]^T\sim \cC\cN(0,Q_{\rm \tx})$. The elements of $Q_{\rm\tx}$ are $q_{\rm\tx}^{ab}$ defined as $q_{\rm\tx}^{ab}=\frac{(\txvd^{(a)})^H\txvd^{(b)}}{\tn}$.

We again apply Varadhan's theorem and Gartner-Ellis theorem\cite{dembo38large} and obtain
\begin{equation*}
    \lim_{\tn\to\infty}\frac{\ln\Xi_{\rm{\tx},n}}{\tn} = \sup_{Q_{\rm \tx}}\inf_{\hat{Q}_{\rm \tx}}\Big[
\ratio G_2(Q_{\rm \tx}) - \sum_{a<b}2q_{\rm \tx}^{ab}\hat{q}_{\rm \tx}^{ab} + G_3(\hat{Q}_{\rm \tx})\Big],
\end{equation*}
where
\begin{equation}
    G_2(Q_{\rm \tx})=\ln\sum_{\rx\in\rxA}\E_{\bw}\prod_{a=0}^{n}\gfunc(\sqrt{\snr\qh}w^{(a)},\rx,\hat{\sigma}^2)
\end{equation}
\begin{equation*}
    G_3(\hat{Q}_{\rm\tx})=\lim_{\tn\to\infty}\frac{1}{\tn}\ln\E_{[\txvd]_0^{n}}\exp\left(\sum_{a<b}2q_{\rm \tx}^{ab}(\txvd^{(a)})^H\txvd^{(b)}\right).
\end{equation*}

Now we apply the RS assumption by considering the off-diagonal elements of $Q_{\rm \tx}$ and $\hat{Q}_{\rm \tx}$ are the same at the saddle point, denoted as $\qx$ and $\qxhat$. Then, we have
\begin{equation*}
    \ratio G_2(Q_{\rm \tx})=-\cG_2(n,\qx),
\end{equation*}
\begin{equation*}
    G_3(\hat{Q}_{\rm \tx})=-\cG_3(n,\qxhat),
\end{equation*}
where $\cG_2(n,\qx)$ and $\cG_3(n,\qxhat)$ are defined in (\ref{eq:G2_def}) and (\ref{eq:G3_def}). Becasue of the symmetry, only real part of $(\txvd^{(a)})^H\txvd^{(b)}$ is needed for computation.

Therefore, (\ref{eq:Entropy_RS_effective}) becomes
\begin{equation}
    \lim_{\tn\to\infty}\frac{1}{\tn}\Ent(\rxvdhat|\hat{\chm})=\frac{1}{\ln2}\cF_2(\qx,\qxhat),
\end{equation}
where $\cF_2(r,\hat{r})$ is defined in (\ref{eq:F2_def}). $(\qx,\qxhat)$ is the saddle point of $\cF_2(r,\hat{r})$, and we have
\begin{equation}
    \frac{\partial \cF_2(r,\hat{r})}{\partial r}=0, \frac{\partial \cF_2(r,\hat{r})}{\partial \hat{r}}=0,
\end{equation}
at $(\qx,\qxhat)$. If there are multiple solutions, we should use the solution that minimize $\cF_2(\qx,\qxhat)$. This proves the rest of the recipe.
\bibliographystyle{IEEEtran}
\bibliography{IEEEabrv,ITA_ref}

\end{document}